\documentclass[oneside,a4paper,12pt]{article}
\usepackage{latexsym}
\usepackage{euscript}
\usepackage{epsfig,amsmath,amssymb}
 
\def\be{\begin{equation}}
\def\ee{\end{equation}}
\def\bea{\begin{eqnarray}}
\def\eea{\end{eqnarray}}

\long\def\symbolfootnote[#1]#2{\begingroup%
\def\thefootnote{\fnsymbol{footnote}}\footnote[#1]{#2}\endgroup} 

 \large\normalsize

\def\Journal#1#2#3#4{{#1} {\bf #2}, #3 (#4)}

\def\PRL{\em Phys. Rev. Lett.}
\def\PRD{{\em Phys. Rev.} D}

\def\be{\begin{equation}}
\def\ee{\end{equation}}
\def\bea{\begin{eqnarray}}
\def\eea{\end{eqnarray}}

\begin{document}
\vspace*{4cm}
\begin{center}

{\Large \bf Cosmology of B-L cosmic strings}

\vspace*{7mm} {\ Rachel Jeannerot}
\vspace*{.25cm}

{\it  ICTP, Strada Costiera 11, P.O. Box 586, 34100 Trieste, Italy}
\vspace*{.25cm}

\end{center}

\begin{abstract}

 ${\rm B\! - \! L}$ cosmic strings form in a wide class of
theories beyond the Standard Model which contain a ${\rm U}(1)_{{\rm
B\! - \! L}}$ gauge symmetry.  They can form at the end of hybrid
inflation and explain, together with inflation, the Cosmic Microwave
Background anisotropies and the formation of large scale
structure. They can produce Cold Dark Matter in the form of the
Lightest SuperParticle and they can be at the origin of the baryon
asymmetry of our universe. One major advantage of these mechanisms is
that they are non-thermal.
\end{abstract}

\section{Motivations}

Theories beyond the Standard Model based on gauge groups G which
contain a ${\rm U}(1)_{{\rm B\! - \! L}}$ gauge symmetry (B and L are
respectively baryon and lepton numbers) are very interesting for both
particle physics and cosmology. First of all, they predict the
existence of right-handed neutrinos and the left-handed neutrinos can
acquire very small masses according to the see-saw mechanism
\cite{seesaw}; the neutrino oscillations discovered by the
SuperKamiokande \cite{superK} are predicted. Next, if the theory is
supersymmetric, the Lightest SuperParticle (LSP) can remain stable
down to low energies and becomes a good Cold Dark Matter (CDM)
candidate. This is due to the fact that ${\rm U}(1)_{{\rm B\! - \!
L}}$ contains a $Z_2$ discrete symmetry which can be left unbroken
down to low energies if Higgs in {\em safe} representations of G are
used to break G down to low energies. This plays the role of R-parity
\cite{Martin} and in such theories the LSP can be automatically
stable. The third main interesting point for cosmology is that
inflation emerges {\em naturally}, i.e. no field nor any symmetry
other than the ones used to build the theory itself are needed for
inflation to arise; and natural values of the parameters are obtain
when constraints from the Cosmic Microwave Background Explorer (COBE)
data are applied \cite{inflationB-L}. If the existence of a unified
gauge group G is assumed, it must somehow be broken down to the
Standard Model gauge group. It can either break directly or via one or
more intermediate gauge symmetry ${\rm G} \rightarrow
...?... \rightarrow {\rm SU}(3)_c \times {\rm SU}(2)_L \times {\rm
U}(1)_Y = {\rm G_{SM}}.$ Therefore, if in the very early Universe
symmetries between particles were described by a gauge group larger
than $G_{SM}$, the Universe must have undergone a series of phase
transitions associated with spontaneous symmetry breakings. These
symmetry breakings may have lead to non-trivial topologies of the
vacuum manifold and topological defects may have formed according to
the Kibble mechanism \cite{Kibble}. It is well known that all unified
theories lead to the formation of topological monopoles which are
topologically stable down to low energies and are in conflict with
observations - this is true as soon as the ${\rm U}(1)_Y$ gauge
symmetry of the Standard Model is embedded in a non-abelian group
which does not contain U(1) factor. Hence some mechanism has to be
invoked to disperse these monopoles. The standard scenario for solving
the monopole problem is inflation. Inflation also solves many of the
cosmological problems such as the horizon problem and predicts the
formation of the large scale structure. Inflation is always confronted
with fine-tuning problems unless supersymmetry is invoked. The
standard scenario for inflation in supersymmetric theories is the
so-called F-term hybrid scenario \cite{Fterm} and it can be
implemented in theories which contain ${\rm U}(1)_{{\rm B\! - \! L}}$
\cite{inflationB-L}. In the simplest model of F-term inflation
topological defects form at the end of inflation, and hence an
intermediate symmetry is needed to solve the monopole problem. In this
case, the general symmetry breaking pattern will thus have to be of
the form ${\rm G} \rightarrow {\rm H} \rightarrow {\rm G_{SM}}$, with
H chosen in such a way that monopoles form when G breaks down to H,
then inflation takes place driven by some scalar field associated with
the breaking of H, and at the end of inflation H spontaneously breaks
down to ${\rm G_{SM}}$ and no unwanted defects must be formed. It is
not easy to find an adequate H. One possibility, is to chose H such
that ${\rm B\! - \! L}$ strings form when H breaks down to ${\rm
G_{SM}}$.

\subsection{${\rm B\! - \! L}$ cosmic strings}

Cosmic strings form according to the Kibble mechanism during the phase
transition associated with the spontaneous symmetry breaking of a
gauge group H down to a subgroup K of H if the first homotopy group of
the vacuum manifold ${\rm H\over K}$ is non trivial. ${\rm B\! - \!
L}$ cosmic strings form when a gauge group H which contains ${\rm
U}(1)_{{\rm B\! - \! L}}$ breaks down to a subgroup K of H which does
not contain ${\rm U}(1)_{{\rm B\! - \! L}}$ if $\pi_1({\rm H\over K})
\neq I$. The Higgs field which forms the string is a Higgs field in a
complex representation of H which breaks H and local ${\rm B\! - \!
L}$ symmetry when acquiring a non-vanishing vacuum expectation value;
we call it $\phi_{{\rm B\! - \! L}}$. In supersymmetric theories, the
superpotential has to be holomorphic and two Higgs superfields
$\Phi_{{\rm B\! - \! L}}$ and $\overline{\Phi}_{{\rm B\! - \! L}}$ in
complex conjugate representations are needed to break ${\rm B\! - \!
L}$; ${\rm B\! - \! L}$ cosmic strings are then made of two Higgs
fields (the scalar components of $\Phi_{{\rm B\! - \! L}}$ and
$\overline{\Phi}_{{\rm B\! - \! L}}$) which wind around the string in
opposite directions. The simplest model in which ${\rm B\! - \! L}$
strings are formed is during the symmetry breaking ${\rm SU}(3)_c
\times {\rm SU(2)}_L \times {\rm U(1)}_R \times {\rm U}(1)_{{\rm B\! -
\! L}} \rightarrow {\rm G_{SM}}$.

\section{${\rm B\! - \! L}$ Strings and Inflation}

\subsection{Hybrid inflation}

As mentioned in the introduction, ${\rm B\! - \! L}$ cosmic strings
can form at the end of F-term hybrid inflation. This is the case when
the superpotential in the sector which breaks H and ${\rm B\! - \! L}$
is given by:
\begin{equation}
W_{{\rm B\! - \! L}} = \alpha S \overline{\Phi}_{{\rm B\! - \! L}}
\Phi_{{\rm B\! - \! L}} - \mu^2 S
\end{equation}
where $S$ is a superfield singlet under H, $\mu$ and $\alpha$ are two
constants which are taken to be positive and $\mu\over \sqrt{\alpha}$
sets the ${\rm B\! - \! L}$ breaking scale.  In such models, the
strings together with inflation generate the cosmological
perturbations, and the strings contribution to the CMB is
non-negligible but model dependent \cite{inflationB-L}. Both the
strings and the inflation perturbations are proportional to $\mu\over
\sqrt{\alpha}$ and COBE data gives the constraint ${\mu\over
\sqrt{\alpha}} \simeq 4.7 \times 10^{15}$ GeV. The mass-per-unit
length of the strings is $\propto {\mu^2 \over \alpha}$. The parameter
$\alpha$ is determined such as to solve the horizon problem and values
of order $10^{-2} - 10^{-3}$ can be found. A realistic model based on
SO(10) has been built \cite{So10}.

Numerical work based on a toy U(1) model has shown that a mixed
scenario with inflation and strings soften the oscillations in the CMB
power spectrum predicted by inflation alone and may well better fit
the data \cite{contaldi}. If this is confirmed by new experimental
data and by improved simulations - using a realistic model and taking
into account the fact that the strings form at the end of inflation
i.e. may have a specific distribution - the motivation for studying
such models would be even stronger.

\subsection{Thermal inflation}

If a superpotential containing only non-renormalisable terms is used
to break ${\rm B\! - \! L}$, such as
\begin{equation}
W_{{\rm B\! - \! L}} = \lambda {(\overline{\Phi}_{{\rm B\! - \! L}} \Phi_{{\rm B\! - \! L}})^2 \over M}
\end{equation}
where $\lambda$ is constant taken to be positive and $M$ is a
superheavy scale beyond which quantum gravity takes place, a period of
thermal inflation takes place \cite{thermal}. This can solve the
monopole problem if the ${\rm B\! - \! L}$ breaking scale is $\sim
10^{11}$ GeV. But it cannot be at the origin of the large scale
structure nor solve the horizon problem. At the end of this low energy
inflationary period fat ${\rm B\! - \! L}$ cosmic strings formed. We
would like to point out that it is not easy to build such a realistic
model.

Thermal inflation dilutes the monopoles previously formed but it also
dilutes any baryon asymmetry which might have been previously
generated; and it is in general very difficult to produce the observed
baryon asymmetry after thermal inflation. However, here, baryogenesis
via leptogenesis can take place at the end of inflation through the
decay of the right-handed neutrinos released by decaying ${\rm B\! -
\! L}$ cosmic string loops \cite{lept}. This scenario is independent of
the reheat temperature of the universe after the inflationary era.

\section{${\rm B\! - \! L}$ cosmic strings produce CDM}

\begin{figure}
\psfig{figure=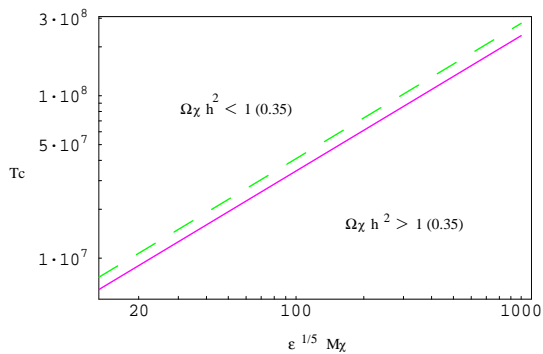,height=2in}
\caption{The critical temperature $T_c$ as a function of the branching
ratio and the LSP mass for $\Omega_\chi h^2 =1$ (solid line) and
$\Omega_\chi h^2 =0.35$ (dashed line). The region above the curves
corresponds to $\Omega_\chi h^2<1$ ($\Omega_\chi h^2<0.35$
respectively) and the region below corresponds to $\Omega_\chi h^2>1$
($\Omega_\chi h^2>0.35$). The latter is excluded by
observations.\label{fig:T}}
\end{figure}

One of the major problem faced by the standard cosmology is to explain
the dark matter component of the Universe. In supersymmetric theories
with conserved R-parity, the LSP is stable and is a good CDM
candidate. As mentioned in the introduction, in theories which predict
${\rm B\! - \! L}$ cosmic strings, the LSP can be automatically stable
\cite{Martin}. When a network of strings is formed, infinite strings
and loops are initially formed. More loops are formed by the
intercommuting of the long strings. String loops rapidly decay by
emitting elementary particles and gravitational radiation. The main
decay channel of ${\rm B\! - \! L}$ cosmic strings loops is into
right-handed neutrinos and sneutrinos. If the LSP is Higgsino, then
these will directly decay into the LSP and if the LSP is bino as is
usually the case in left-right models, subsequent decay will produce
the LSP. A scaling network of ${\rm B\! - \! L}$ cosmic strings can
thus produce LSPs and these in non-negligible quantities
\cite{LSP}. This mechanism of LSP production, as is the leptogenesis
mechanism, is non-thermal.  We find that the LSP number density today
released by decaying cosmic string loops is given by:
\begin{equation}
Y^{nonth}_{LSP} = {n^{nonth}_{LSP}\over s} = {{6.75} \over {\pi}} \epsilon \nu
\lambda^2 \Gamma_{loops}^{-2} g_{*_{T_{\chi}}}^{-9 \over 4}
g_{*_{T_{\chi}}}^{3 \over 4} \,  
 M_{pl}^2\, {T_{\chi}^4 \over T_{{\rm B\! - \! L}}^6} \, , \label{eq:Ynonth}
\end{equation}
where $\epsilon$ denotes the branching ratio of the right-handed
neutrinos into LSP, $\lambda$ is the the Higgs self coupling and
$\Gamma_{loops} \simeq 10-20$ is a numerical string parameter.
$g_{*_{T_{{\rm B\! - \! L}}}}$ and $g_{*_{T_{\chi}}}$ count the number
of massless degrees of freedom at the critical temperature $T_c
=T_{{\rm B\! - \! L}}$ at which the strings form and at the LSP
freeze-out temperature $T_{\chi}$ respectively. $M_{pl}$ is the Planck
mass. In Fig.\ref{fig:T} we plot $T_c$ a
function of the LSP mass $M_\chi$. The region above each curves
corresponds to $\Omega_\chi h^2<1$ ($\Omega_\chi h^2 < 0.35$
respectively), and the region below to $\Omega_\chi h^2 >1$
($\Omega_\chi h^2 >0.35$ respectively); this region is excluded by
observations.

\section{Conclusions}

${\rm B\! - \! L}$ cosmic strings can form in all unified theories
based on gauge groups with rank greater than five. They are very
interesting cosmologically because they can explain the
matter-antimatter asymmetry of the Universe, they can form at the end
of inflation and modify the CMB power spectrum, and they can produce
CDM in non-negligible quantities.

\section*{Acknowledgments}

The Author would like to thank her collaborators R. Brandenberger
and X. Zhang.

\end{document}